\documentclass[doublecol]{epl2}

\usepackage{amsmath}
\usepackage{amssymb}

\title{Blow-up solitons at the nonlinear stage of the two-stream instability in quantum plasmas}

\author{V. M. Lashkin \inst{1}}

\institute{
 \inst{1} Institute for Nuclear Research - Pr. Nauki
47, Kyiv 03680, Ukraine}


\abstract{The nonlinear evolution of the quantum two-stream
instability in a plasma with counter-streaming electron beams is
studied. It is shown that in the long-wave limit the nonlinear
stage of the instability can be described by the elliptic
nonlinear string equation. We present two types of the nonlinear
solutions. The first one is an unstable nonlinear mode that is
continuously related with the growing linear solution and the
second one is a pulsating soliton. We show that both of these
solutions blow up in a finite time.}

\begin{document}

\maketitle

\section{Introduction}
The two-stream instability is one of the most known instability in
plasma physics \cite{Trievel86}. This instability may appear when
the plasma particles stream with different velocities. In
particular, this may be the classical instability of an electron
beam in a plasma \cite{Pierce48,Bohm49} and the Buneman
instability \cite{Buneman59} which are widespread in both
laboratory and space plasmas. In the context of quantum plasmas,
the problem of the two-stream instability was first studied by
Haas \textit{et al.} \cite{Haas2000}. Quantum effects in plasmas
are important in the limit of low-plasma temperature and
high-particle number density \cite{Melrose2008,HaasBook2011}. Such
plasmas are ubiquitous in microelectronic devices, in dense
astrophysical plasma, in microplasmas, and in laser plasmas (for
example, see the reviews \cite{Shukla2011,Vladimirov2011} and
references therein). There are two well-known models for
describing the quantum effects in a plasma. The Wigner and Hartree
models are based upon the Wigner-Poisson and
Schr\"{o}dinger-Poisson systems which, respectively, correspond to
the statistical and hydrodynamic description of the plasma
particles. Authors \cite{Haas2000} have introduced a quantum
multistream model by using the nonlinear Schr\"{o}dinger-Poisson
system. In particular, they derived the dispersion relation for
the quantum two-stream instability and showed the existence of
new, purely quantum, unstable branches which do not appear in
classical plasmas. Later on, the quantum hydrodynamic description
of the quantum two-stream instability \cite{Haas2000,Haas2001} was
extended to the kinetic case \cite{Anderson2002} which was based
on the quantum mechanical Wigner formalism. Since then, the
quantum two-stream instability has attracted considerable
attention and been discussed by many authors for different plasma
regimes such as dusty quantum plasma \cite{Ali2007},
electron-positron-ion quantum plasma \cite{Mushtaq2008}, dense
electron-ion plasma \cite{Son2014}, relativistic plasma
\cite{Haas2012}, and electron-hole quantum semiconductor plasma
\cite{Zeba2012}. The effects of the exchange interaction
\cite{Kull2017} and the electron spin \cite{Hussain2019} for the
quantum two-stream instability have been also investigated.
Necessary validity conditions for the cold quantum two-stream
model were discussed in ref. \cite{Haas2009}. Recently, it was
also shown \cite{Brodin2019} that the hydrodynamic model gives a
decent approximation of the exchange effects for sufficiently long
wavelengths.

A nonlinear stage of the quantum two-stream instability was
studied back in the original work \cite{Haas2000} by numerical
simulation, where, in particular, spatially periodic, stationary
states which  survive over long times were predicted. The aim of
this Letter is to derive a nonlinear evolutional equation
describing the nonlinear stage of the quantum two-stream
instability with cold electrons in the long-wavelength limit and
to present corresponding exact analytical solutions. To our
knowledge, this equation in the form of the elliptic nonlinear
string (ENS) equation in this context was not previously obtained
even for classical plasmas. In the classical plasma, a nonlinear
stage of the long-wavelength Buneman instability was studied in
the model of the Boussinesq equation, where the dispersion
corresponds to warm electrons \cite{Kono1990}.  The ENS equation
differs from the classical Boussinesq equation in the sign before
a fourth derivative term. Note that when considering the nonlinear
stage of the quantum two-stream instability in
electron-positron-ion plasmas the Korteweg-de Vries (KdV) equation
was obtained by employing the reductive perturbation technique
\cite{Mushtaq2008}. The ENS equation as well as the KdV equation
is completely integrable and admits $N$-soliton solution
\cite{Zakharov2002}. A significant difference, however, is that
the dispersion relation of the linearized KdV equation does not
corresponds to any instability at all, while the dispersion
relation of the ENS equation completely coincides with the one of
the two-stream instability in the long-wavelength limit.

\section{Model equation}
The longitudinal dielectric response function of a cold plasma
with the "quantum recoil" term for two counter propagating
electron beams with equal equilibrium number densities and
velocities has the form \cite{Haas2000,HaasBook2011}
\begin{gather}
\varepsilon(\omega,k)=1-\frac{\omega_{p}^{2}/2}
{(\omega-kv_{0})^{2}-\hbar^{2}k^{4}/(4m^{2})} \nonumber \\
-\frac{\omega_{p}^{2}/2}{(\omega+kv_{0})^{2}-\hbar^{2}k^{4}/(4m^{2})},
\label{linear_disp}
\end{gather}
where $\omega$ and $k$ are the frequency and the wave number
respectively, $\omega_{p}=\sqrt{4\pi e^{2} n_{0}/m}$ is the
electron plasma frequency, $n_{0}$ is the equilibrium plasma
density, $e$ is the elementary charge, $m$ is the electron mass,
$\hbar$ is the Planck constant divided by $2\pi$, and $v_{0}$ is
the stream velocity. Charge neutrality is provided by the
motionless background ions. The solution of eq.
(\ref{linear_disp}) for $\omega^{2}$ has two branches
\cite{Haas2000}
\begin{equation}
\label{linear_disp2}
\omega^{2}_{\pm}=\frac{\omega_{p}^{2}}{2}+k^{2}v_{0}^{2}+\frac{\hbar^{2}k^{4}}{4m^{2}}\pm
\frac{\omega_{p}^{2}}{2}\sqrt{1+\frac{8k^{2}v_{0}^{2}}{\omega_{p}^{2}}+
\frac{4\hbar^{2}k^{6}v_{0}^{2}}{m^{2}\omega_{p}^{4}}}.
\end{equation}
The solution with the sign "$+$" is always positive and gives
stable oscillations. The other solution can be negative
($\omega^{2}<0$) and the instability occurs provided that
\cite{Haas2000}
\begin{equation}
\label{instability_condition}
[H^{2}K^{2}-4][H^{2}K^{4}-4K^{2}+4]<0,
\end{equation}
where the rescaled variables are
\begin{equation}
H=\hbar\omega_{p}/(mv_{0}^{2}), \, \, K=kv_{0}/\omega_{p}.
\end{equation}
In this Letter, we restrict to the case of the long-wave limit
when $K^{2}\ll 1$ and $H^{2}K^{4}\ll 1$. Note that, generally
speaking, a strongly quantum regime $H^{2}\gg 1$ can also take
place. Then the dispersion relation (\ref{linear_disp2}) with the
sign "$-$" takes the form
\begin{equation}
\label{linear_disp3}
\omega^{2}=-k^{2}v_{0}^{2}+\frac{4v_{0}^{4}}{\omega_{p}^{2}}\left(1+\frac{H^{2}}{16}\right)
k^{4}
\end{equation}
and predicts the instability with the growth rate
\begin{equation}
\label{growth_rate} \gamma=
kv_{0}\sqrt{1-4(1+H^{2}/16)k^{2}v^{2}_{0}/\omega^{2}_{p}}.
\end{equation}
We address the nonlinear stage of this instability.

In the following we use the notation
\begin{equation}
\sum_{q=q_{1}+q_{2}} \cdots \rightarrow\int \cdots \,\delta
(q-q_{1}-q_{2})\frac{dq_{1}}{(2\pi)^{2}}\frac{dq_{2}}{(2\pi)^{2}},
\end{equation}
where $\delta (x)$ is the Dirac delta function and $q=(\omega,k)$.
In the one-dimensional space the Wigner kinetic equation
\cite{Moyal1949,Klimontovich1952,Tatarskii1983} for the quantum
electron distribution function (Wigner function) $F(x,v,t)$ can be
written as
\begin{gather}
\frac{\partial F}{\partial t}+v\frac{\partial F}{\partial
x}=-\frac{iem}{2\pi\hbar^{2}}\int\int d\lambda dv^{'}\exp
\left[i\frac{m}{\hbar}(v-v^{'})\lambda\right]  \nonumber \\
\times\left[\varphi\left(x+\frac{\lambda}{2},t\right)
-\varphi\left(x-\frac{\lambda}{2},t\right)\right]F(x,v^{'},t),
\label{basic_tatar}
\end{gather}
where $\varphi$ is the electrostatic potential. In the momentum
space eq. (\ref{basic_tatar}) can be written as
\begin{gather}
(\omega-kv) f_{q}(v)=\frac{em}{2\pi\hbar^{2}}\int\int d\lambda
dv^{'}\exp
\left[i\frac{m}{\hbar}(v-v^{'})\lambda\right]  \nonumber \\
\times\left[\left(\mathrm{e}^{ik\lambda/2}-\mathrm{e}^{-ik\lambda/2}\right)\varphi_{q}f^{(0)}(v^{'})
 \right. \nonumber \\ \left. +\sum_{q=q_{1}+q_{2}}\left(\mathrm{e}^{ik_{1}\lambda/2}-\mathrm{e}^{-ik_{1}\lambda/2}\right)
\varphi_{q_{1}}f_{q_{2}}(v^{'})\right] \label{basic_kinetic1}
\end{gather}
where $f_{q}(v)$ is the deviation of the electron distribution
function of each stream from the equilibrium one $f^{(0)}(v)$, and
$\varphi$ is the electrostatic potential. The distribution
function $f^{(0)}(v)$ is normalized to the equilibrium plasma
density of each stream, $\int f^{(0)}(v)dv=n_{0}/2$. Integrating
over $\lambda$ in eq. (\ref{basic_kinetic1}) yields
\begin{equation}
\exp \left[i\frac{m}{\hbar}(v-v^{'})\pm\frac{ik}{2}\right]\lambda
\rightarrow 2\pi\delta
\left[\frac{m}{\hbar}(v-v^{'})\pm\frac{k}{2}\right],
\end{equation}
and then integrating over $v^{'}$ yields
\begin{gather}
(\omega-kv)
f_{q}(v)=\frac{e}{\hbar}\varphi_{q}\left[f^{(0)}\left(v+\frac{\hbar
k}{2m}\right)-f^{(0)}\left(v-\frac{\hbar k}{2m}\right)\right]
\nonumber \\
+\frac{e}{\hbar}\sum_{q=q_{1}+q_{2}}\varphi_{q_{1}}\left[f_{q_{2}}\left(v+\frac{\hbar
k_{1}}{2m}\right)-f_{q_{2}}\left(v-\frac{\hbar
k_{1}}{2m}\right)\right].
 \label{basic_kinetic}
\end{gather}
We present the function $f_{q}(v)$ as a series in powers of the
field strength (i. e. $f_{q}^{(n)}\sim \varphi^{n}$)
\begin{equation}
\label{series} f_{q}(v)=\sum_{n=1}^{\infty}f_{q}^{(n)}(v).
\end{equation}
In the linear approximation from eqs. (\ref{basic_kinetic}) and
(\ref{series}) we have
\begin{equation}
f_{q}^{(1)}=\frac{e\varphi_{q}}{\hbar
(\omega-kv)}\left[f^{(0)}\left(v+\frac{\hbar
k}{2m}\right)-f^{(0)}\left(v-\frac{\hbar k}{2m}\right)\right],
\end{equation}
and then one can write the recurrence relation
\begin{gather}
f_{q}^{(n)}=\frac{e}{\hbar
(\omega-kv)}\sum_{q=q_{1}+q_{2}}\varphi_{q_{1}}\left[f_{q_{2}}^{(n-1)}\left(v+\frac{\hbar
k_{1}}{2m}\right) \right. \nonumber \\
 \left. -f_{q_{2}}^{(n-1)}\left(v-\frac{\hbar
k_{1}}{2m}\right)\right]. \label{recur1}
\end{gather}
For the nonlinear terms ($n \geqslant 2$) we use the
quasiclassical approximation and in the limit $\hbar k/(2m)\ll v$
in eq. (\ref{recur1})  one can expand
\begin{equation}
f_{q_{2}}^{(n-1)}\left(v\pm\frac{\hbar k_{1}}{2m}\right)\approx
f_{q_{2}}^{(n-1)}(v)\pm\frac{\partial f_{q_{2}}^{(n-1)}}{\partial
v}\frac{\hbar k_{1}}{2m}
\end{equation}
whence we get
\begin{equation}
f_{q}^{(n)}=\frac{e}{m(\omega-kv)}
\sum_{q=q_{1}+q_{2}}k_{1}\varphi_{q_{1}}\frac{\partial
f_{q_{2}}^{(n-1)}}{\partial v}. \label{recur2}
\end{equation}
Retaining terms in eq. (\ref{series}) up to second order in the
wave fields and substituting $f_{q}$ into the Poisson equation
\begin{equation}
k^{2}\varphi_{q}=-4\pi e \sum_{\alpha}\int f_{q}(v)dv ,
\end{equation}
where $\sum$ stands for summation over the different electron
streams labeled by $\alpha=+,-$, and the ion contribution is
omitted, we get
\begin{equation}
\label{nonlin_eq1} \varepsilon_{q}\varphi_{q}=\sum_{q=q_{1}+q_{2}}
V_{q_{1},q_{2}}\varphi _{q_{1}}\varphi_{q_{2}},
\end{equation}
where
\begin{equation}
\label{linear_responce} \varepsilon_{q}=1+\frac{4\pi e^{2}}{\hbar
k^{2}}\sum_{\alpha}\int\frac{\left[f^{(0)}\left(v+\frac{\hbar
k}{2m}\right)-f^{(0)}\left(v-\frac{\hbar
k}{2m}\right)\right]}{(\omega-kv)}dv
\end{equation}
is the linear electron dielectric response function, and the
interaction matrix element $V_{q_{1},q_{2}}$ is determined by
\begin{gather}
V_{q_{1},q_{2}}=\frac{e}{2m}
\frac{\omega^{2}_{p}}{n_{0}k^{2}}\sum_{\alpha}\int
\frac{k_{1}}{[(\omega_{1}+\omega_{2})-(k_{1}+k_{2})
v]} \nonumber \\
\times\frac{\partial}{\partial
v}\frac{k_{2}}{(\omega_{2}-k_{2}v)}\frac{\partial
f^{(0)}_{\alpha}}{\partial v}dv +(\omega_{1},k_{1}
\rightleftarrows \omega_{2},k_{2}). \label{matrix1}
\end{gather}
Note that the expression (\ref{matrix1}) for the interaction
matrix element $V_{q_{1},q_{2}}$ is written in a symmetrized form.
Singularities in the denominators in eqs. (\ref{linear_responce})
and (\ref{matrix1})  are avoided, as usual, using Landau's rule by
replacing  $\omega\rightarrow\omega+i0$ and then
\begin{equation}
(\omega-kv)^{-1}=\mathcal{P}(\omega-kv)^{-1}-i\pi\delta
(\omega-kv), \label{Landau_lin}
\end{equation}
\begin{gather}
[\omega_{1}+\omega_{2}-(k_{1}+k_{2})v]^{-1}=\mathcal{P}[\omega_{1}+\omega_{2}-(k_{1}+k_{2})v]^{-1}
\nonumber \\ -i\pi\delta [\omega_{1}+\omega_{2}-(k_{1}+k_{2})v],
\label{Landau_nonlin}
\end{gather}
where $\mathcal{P}$ is the principal value of the integrals.
Imaginary parts in eqs. (\ref{Landau_lin}) and
(\ref{Landau_nonlin}) account for the linear and nonlinear Landau
damping respectively. In this paper we do not take into account
the damping and only the principal value of the integrals is
understood. After two partial integrations in eq. (\ref{matrix1})
one can write
\begin{gather}
V_{q_{1},q_{2}}=\frac{e}{2m}
\frac{\omega^{2}_{p}}{n_{0}k^{2}}\sum_{\alpha}\int \left[
\frac{2k^{2}k_{1}k_{2}}{(\omega-kv)^{3} (\omega_{2}-k_{2}v)}
\right.
 \nonumber \\
 \left. +\frac{kk_{1}k_{2}^{2}}{(\omega-
 kv)^{2}(\omega_{2}-k_{2}v)^{2}}\right]f^{(0)}_{\alpha}
dv+(\omega_{1},k_{1}\rightleftarrows\omega_{2},k_{2}),
\label{matrix2}
\end{gather}
where $\omega=\omega_{1}+\omega_{2}$ and $k=k_{1}+k_{2}$.
Expanding $\varepsilon(\omega,k)$ given by eq. (\ref{linear_disp})
near the $\omega_{k}^{2}$ determined by eq. (\ref{linear_disp3})
yields
\begin{equation}
\label{expansion} \varepsilon(\omega,k)=\varepsilon(\omega_{k},k)+
\varepsilon^{'}(\omega_{k}^{2})(\omega^{2}-\omega_{k}^{2})
\end{equation}
where $\varepsilon^{'}(\omega_{k}^{2})\equiv\partial \varepsilon
(\omega^{2})/\partial\omega\mid_{\omega^{2}=\omega_{k}^{2}}$ and
in the leading order one can get
\begin{equation}
\label{epsilon_der_}
\varepsilon^{'}(\omega_{k}^{2})=-\frac{\omega_{p}^{2}}{4k^{4}v_{0}^{4}}.
\end{equation}
After substituting eq. (\ref{expansion}) into eq.
(\ref{nonlin_eq1})  we have
\begin{equation}
\label{nonlin_eq_basic}
(\omega^{2}-\omega_{k}^{2})\varphi_{q}=\frac{1}{\varepsilon^{'}(\omega_{k}^{2})}
\sum_{q=q_{1}+q_{2}} V_{q_{1},q_{2}}\varphi
_{q_{1}}\varphi_{q_{2}}.
\end{equation}
Considering the monoenergetic electron streams, $v_{0}\gg v_{Te}$,
where $v_{Te}$ is the thermal electron velocity, one can write for
the full electron distribution function $f^{(0)}$
\begin{equation}
\label{distribution}
f^{(0)}=f^{(0)}_{-}+f^{(0)}_{+}=\frac{n_{0}}{2}[\delta
(v-v_{0})+\delta (v+v_{0})].
\end{equation}
Substituting eq. (\ref{distribution}) into eq.
(\ref{linear_responce})  and then after suitable changes of
variables one can obtain the dielectric function
(\ref{linear_disp}). Substituting eq. (\ref{distribution}) into
eq. (\ref{matrix2}) we obtain
\begin{gather}
V_{k_{1},k_{2}}=\frac{e}{2m}\frac{\omega_{p}^{2}}{k^{2}}\left[2k^{2}k_{1}k_{2}
\left(\frac{1}{\Omega^{3}_{-}\Omega_{-,2}}+\frac{1}{\Omega^{3}_{+}\Omega_{+,2}}\right)\right.
\nonumber \\
\left.
+kk_{1}k_{2}^{2}\left(\frac{1}{\Omega^{2}_{-}\Omega_{-,2}^{2}}
+\frac{1}{\Omega^{2}_{+}\Omega_{+,2}^{2}}\right)\right]
+(\omega_{1},k_{1}\rightleftarrows \omega_{2},k_{2}),
\label{matrix3}
\end{gather}
where $\Omega_{\pm}=\omega\pm kv_{0}$,
$\Omega_{\pm,2}=\omega_{2}\pm k_{2}v_{0}$. The wave dispersion in
eq.  (\ref{linear_disp3}) has an acoustic type and in the leading
term satisfies the three-wave resonance condition
\begin{equation}
\label{resonance} \omega_{k}=\omega_{k_{1}}+\omega_{k_{2}}, \, \,
k=k_{1}+k_{2}.
\end{equation}
Taking into account eq. (\ref{linear_disp3}) and eq.
(\ref{resonance}) when calculating (\ref{matrix3}), and then
substituting eq. (\ref{epsilon_der_}) and eq. (\ref{matrix3}) into
(\ref{nonlin_eq_basic}) we finally obtain
\begin{gather}
\label{Boussinesq0}
\left[\omega^{2}+k^{2}v_{0}^{2}-\frac{4v_{0}^{4}}{\omega_{p}^{2}}\left(1
+\frac{H^{2}}{16}\right)k^{4}\right]\varphi_{q}= \nonumber \\
-\frac{3e}{m}k^{2} \sum_{q=q_{1}+q_{2}}\varphi
_{q_{1}}\varphi_{q_{2}}.
\end{gather}
In the physical space this equation takes the form
\begin{equation}
\label{Boussinesq1} \frac{\partial^{2}\varphi}{\partial
t^{2}}+v_{0}^{2}\frac{\partial^{2}\varphi}{\partial x^{2}}
+\frac{4v_{0}^{4}}{\omega_{p}^{2}}\left(1+\frac{H^{2}}{16}\right)\frac{\partial^{4}\varphi}{\partial
x^{4}}+\frac{3e}{m}\frac{\partial^{2}\varphi^{2}}{\partial
x^{2}}=0.
\end{equation}
In the linear approximation, taking $\varphi\sim \exp
(-ikx+i\omega t)$,  the equation (\ref{Boussinesq1}) yields the
dispersion relation (\ref{linear_disp3}). After rescaling
\begin{equation}
t\rightarrow \omega_{p}t,\,\,\, x\rightarrow
\frac{\omega_{p}}{v_{0}}x, \,\,\, \varphi\rightarrow
\frac{e\varphi}{8mv_{0}^{2}(1+H^{2}/16)},
\end{equation}
equation (\ref{Boussinesq1}) can be written in the following
dimensionless form
\begin{equation}
\label{Boussinesq2}
\varphi_{tt}+\varphi_{xx}+4\left(1+\frac{H^{2}}{16}\right)(\varphi_{xx}+6\varphi^{2})_{xx}=0.
\end{equation}
The linear part of this equation corresponds to the elliptic
operator. The sign before the nonlinear term is not essential and
can be changed by replacing $\varphi\rightarrow -\varphi$. The
equation (\ref{Boussinesq2}) with negative signs before
$\varphi_{xx}$ and $\varphi_{xxxx}$ is the classical Boussinesq
equation and its $N$-soliton solutions were obtained by Hirota
\cite{Hirota73} using his bilinearization method. The linear
dispersion relation of the Boussinesq equation predicts the
instability for short waves unlike the equation of a nonlinear
string (\ref{Boussinesq2}) describing the long-wave instability.
Yadjima pointed out \cite{Yadjima83} that in the Hirota's
solutions also implicitly present the solutions which are
continuously related to the unstable solution of the corresponding
linear equation. The solution in \cite{Yadjima83} describes the
nonlinear stage of the linear instability. Under this, the
over-stabilization takes place the unstable mode neither grows
unlimitedly nor saturates at some level but takes the maximum and
thereafter damps. As we see below, the situation is essentially
different for eq. (\ref{Boussinesq2}).

\section{Blow-up nonlinear solutions}
The equation (\ref{Boussinesq2}) with negative sign only before
$\varphi_{xx}$ is known as the equation of nonlinear string.
Although this equation is linearly stable, Kalantarov and
Ladyzhenskaya \cite{Ladyzhenskaya78} were apparently the first to
show that real solutions with initial data satisfying the
inequality $\mathcal{H}<0$, where $\mathcal{H}$ is the
hamiltonian, blow up in a finite time. For the same equation the
possibility of collapsing solitons with $\mathcal{H}>0$ was shown
in ref. \cite{Turitsyn83}. Formation of a singularity (collapse of
solitons) and decay of solitons was demonstrated in ref.
\cite{Zakharov2002}, where the continuous (nonsolitonic) spectrum
was also investigated.

The recipe suggested in ref. \cite{Yadjima83} can be extended to
the case of the elliptic equation of a nonlinear string
(\ref{Boussinesq2}) describing the long-wave instability. Then one
can find an exact solution of eq. (\ref{Boussinesq2}) in the form
\begin{equation}
\label{exact_solution1} \varphi=k^{2}\frac{\{\sqrt{P}\cosh[\gamma
t+\ln(a\sqrt{P}/2k^{2})]\cos kx -1\}}{\{\sqrt{P}\cosh[\gamma
t+\ln(a\sqrt{P}/2k^{2})]-\cos kx\}^{2}}
\end{equation}
where
\begin{equation}
\label{k_K} \gamma=k\sqrt{1-4(1+H^{2}/16)k^{2}},\,\,
P=\frac{1-16(1+H^{2}/16)k^{2}}{1-4(1+H^{2}/16)k^{2}}.
\end{equation}
Here, $a$ and $k$ are arbitrary parameters satisfying the
conditions $a>0$ and $4(1+H^{2}/16)k^{2}<1$ so that $P>0$. By
expanding eq. (\ref{exact_solution1}) in $a$ one can obtain
\begin{gather}
\label{expansion_linear} \varphi=a\exp (\gamma t)\cos
kx+\frac{a^{2}}{k^{2}}\exp (2\gamma t)\cos 2kx \nonumber \\
+\frac{3a^{3}}{4k^{4}}\exp (3\gamma t)\left[\cos
3kx-\frac{k^{4}}{\gamma^{2}}\cos kx\right]+\dots ,
\end{gather}
which is in agreement to every order in $a$ with the expression
obtained from the perturbation theory. At the same time,
neglecting the nonlinear term in eq. (\ref{Boussinesq2}) yields
the growing solution
\begin{equation}
\label{linear_growth} \varphi=a\exp (\gamma t)\cos kx
\end{equation}
with the growth rate $\gamma$ given by eq. (\ref{growth_rate}) and
corresponds to the linear dispersion relation
(\ref{linear_disp3}). Thus, the solution (\ref{exact_solution1})
is continuously related to the linear solution
(\ref{linear_growth}). Since $P<1$ it is seen that the solution
will blow up and becomes infinite at time
\begin{equation}
t_{cr}=\frac{1}{\gamma}\ln\frac{2k^{2}}{a(1+\sqrt{1-P})}.
\end{equation}
This is in contrast to the behavior predicted in ref.
\cite{Yadjima83} for the Boussinesq equation, when the unstable
solution initially grows with time then reaches the maximum at
some time and thereafter damps so that the over-saturating of the
instability takes place.

Another exact solution of eq. (\ref{Boussinesq2}) is the pulsating
(breather) soliton. One can obtain this solution just putting
$\gamma=i\Omega$, where $\Omega$ is real, and $a=2k^{2}/\sqrt{P}$
in eq. (\ref{exact_solution1}), so that there is only one
arbitrary parameter $q$ in the solution
\begin{equation}
\label{exact_solution2} \varphi=q^{2}\frac{(1+\sqrt{Q}\cos\Omega
t\cosh qx )}{(\sqrt{Q}\cos\Omega t+\cosh qx )^{2}},
\end{equation}
where
\begin{equation}
\label{q_Q} \Omega=q\sqrt{1+4(1+H^{2}/16)q^{2}},\,
Q=\frac{1+16(1+H^{2}/16)q^{2}}{1+4(1+H^{2}/16)q^{2}}.
\end{equation}
It is seen that $Q>1$ and the solution (\ref{exact_solution2}) has
a singularity at some time. The explosion time is
\begin{equation}
t_{cr}=\frac{1}{\Omega}\arccos\left(-\frac{1}{\sqrt{Q}}\right).
\end{equation}
It is seen that in the strongly quantum regime $H^{2}\gg 1$ the
amplitudes, characteristic sizes and times of the found nonlinear
structures are completely determined by the quantum effect. In the
limit $H^{2}\rightarrow 0$, we  readily obtain the case of
classical plasma, which for the two cold  countering electron
streams was also not discussed earlier in the literature in this
context.

A singularity in the solutions indicates that the model
(\ref{Boussinesq1}) is no longer valid and the full model
\cite{Haas2000} should be used. In reality, the singularity can
never be reached and the blow-up is prevented by including extra
effects such as Landau damping, higher-order dispersion and
nonlinearities. In our case of the monoenergetic electron streams,
the stabilization can occur long before thermalization of the
beams or reaching the characteristic scale length of the order of
the Debye radius due to the electron trapping in wave
electrostatic potential well when $e\varphi\sim mv_{0}^{2}$
\cite{Rukhadze84} which is in agreement with \cite{Haas2000}.

\section{Conclusions}
We have derived the nonlinear equation describing the development
of the quantum two-stream instability in cold plasmas with counter
streaming electrons. This equation has a form of the elliptic
nonlinear string equation. We have presented two type of the exact
solutions of this equation. The first one is an unstable nonlinear
mode that continuously related with the growing linear solution
and the second one is a pulsating soliton. We have shown that both
of these solutions blow up in a finite time.

\end{document}